\documentclass{PoS}

\PoS{PoS(LAT2005)147}

\usepackage{epsfig}

\title{Chiral limit of 2-color QCD at strong couplings }

\ShortTitle{2-color QCD}

\author{Shailesh Chandrasekharan and \speaker{Fu-Jiun Jiang}
\thanks{This article contains the combined contents of 
  the poster by Chandrasekharan and the talk by Jiang.}\\
        Box 90305, Duke University, Durham NC 27708, USA\\
        E-mail: \email{sch@phy.duke.edu}}

\abstract{
We study two-color lattice QCD with massless staggered fermions in 
the strong coupling limit using a new and efficient cluster algorithm. 
We focus on the phase diagram of the model as a function of temperature 
$T$ and baryon chemical potential $\mu$  by working on $L_t \times L^d$ 
lattices in both $d=2,3$. In $d=3$ we find that at $\mu=0$ the ground 
state of the system breaks the global $U(2)$ symmetry present in the model 
to $U(1)$, while the finite temperature phase transition (with $L_t=4$) 
which restores the symmetry is a weak first order transition. In $d=2$
we find evidence for a novel phase transition similar to the 
Berezinky-Kosterlitz-Thouless phenomena. On the other hand the 
quantum ($T=0$) phase 
transition to a symmetric phase as a function of $\mu$ is second order 
in both $d=2,3$ and belongs to the mean field universality class.}

\FullConference{XXIIIrd International Symposium on Lattice Field Theory\\
                 25-30 July 2005\\
                 Trinity College, Dublin, Ireland}

\begin{document}

\section{INTRODUCTION}

Two-color QCD has been extensively studied over the years both 
theoretically \cite{Kogut:1999iv,Wirstam:1999ds,Ratti:2004ra,Lenaghan:2001sd} 
and numerically \cite{Kogut:1985un,Kogut:1987ev,Kaczmarek:1998wr,Hands:1999md,Hands:2000ei,Kogut:2001na,Kogut:2003ju,Skullerud:2003yc}. Although a lot 
of progress has been made in uncovering important qualitative features of 
this theory, many interesting quantitative questions remain:
\begin{enumerate}
\item What is the order of the finite temperature chiral transition at zero
and non-zero chemical potentials? 
\item Can the low energy physics at small $T$ and $\mu$ be captured by 
chiral perturbation theory? An answer to this question was attempted in 
\cite{Kogut:2003ju}.
\item What is the order of the phase transition that occurs when the
lattice gets saturated with baryons at $T=0$?
\item What is the phase structure in two spatial dimensions since spontaneous 
symmetry breaking is forbidden at finite temperatures. 
\end{enumerate}
The reason for the lack of quantitative progress can be traced to the fact 
that all previous studies have been limited to small lattice sizes and 
relatively large quark masses; a problem which haunts all numerical studies 
of strongly correlated fermionic systems. 

In this work we try to make progress by considering the strong coupling
limit. Although this limit has the worst lattice artifacts, it contains 
some of the essential physics, namely confinement and chiral symmetry 
breaking. On the other hand recent advances in Monte Carlo algorithms 
allow us to study the chiral limit on large lattices with
relative ease in the strong coupling limit \cite{Adams:2003cc}.
In this work we extend these algorithms and apply it to study
strong coupling two-color QCD with staggered fermions. This theory is 
especially interesting due to an enhanced $U(2)$ symmetry 
at zero quark mass and baryon chemical potential. It was originally 
considered in \cite{Dagotto:1986gw,Klatke:1989xy} and was recently 
reviewed in \cite{Nishida:2003uj}. However, many of the questions 
raised above remain unanswered even in this simplified limit.

\section{THE MODEL}
\label{model}

Our model lives on a $d+1$ dimensional hyper-cubic lattice with sites
$x\equiv(x_t;x_1,x_2,..,x_d)$. The size of the lattice is taken to be
$L_t\times L^d$ and is periodic in all directions. The action of our 
model is
\begin{equation}
S=-\sum_{x,\alpha}r_\alpha \eta_{\alpha}(x)
\Bigg[
e^{\mu a_t \delta_{t,\alpha}}
\overline{\chi}(x)U_{\alpha}(x)\chi(x+\hat{\alpha})
-e^{-\mu a_t \delta_{t,\alpha}}
\overline{\chi}(x+\hat{\alpha})U_{\alpha}^{\dagger}(x)\chi(x)\Bigg].
\label{scact}
\end{equation}
The Grassmann fields $\overline{\chi}(x)$ and $\chi(x)$ represent row and 
column vectors with $2$ color components associated to the lattice site $x$.
The color component of the quark fields will be denoted as $\chi_a,a=1,2$. 
The gauge fields $U_{\alpha}(x)$ are elements of $SU(2)$ group and live on 
the links between $x$ and $x+\hat{\alpha}$ where $\alpha=1,2,..,d$ for
spatial links and $\alpha = t$ for temporal link. The factor $r_\alpha=1$ 
for $\alpha=1,2,..,d$ and $r_t = \frac{1}{a_{t}}$ with
$a_{t}$ being the asymmetry factor between spatial and temporal 
lattice spacing. This asymmetry allows us to study finite temperature 
behavior \cite{Boyd:1991fb}. 

 A discussion of the relevant symmetries of the action (\ref{scact}) can be 
found in \cite{Hands:1999md,Nishida:2003uj}. As explained in these references
when $\mu=0$ our model has a $U(2)$ global symmetry:
\begin{equation}
X_{o}\rightarrow VX_{o},\quad
\overline{X}_{e}\rightarrow\overline{X}_{e}V^{\dagger},\qquad 
V = \exp(i\vec{\alpha}\cdot\vec{\sigma}+i\phi) \in U(2).
\label{u2symm}
\end{equation}
where $\overline{X}_{e}$ and $X_{o}$ are given by
\begin{equation}
\overline{X}_{e}=(\overline{\chi}_{e},-\chi_{e}^{tr}\tau_{2}),
\quad\quad X_{o}=\left(\begin{array}{c}
\chi_{o}\cr -\tau_{2}\overline{\chi}_{o}^{tr}\end{array}\right)
\label{twocompdef}
\end{equation}
and the subscripts $e$ and $o$ refer to even and odd sites. Note in our 
notation $\vec{\sigma}$ are Pauli matrices that mix $\chi$ and 
$\overline{\chi}^{tr}$ present in $X_o$ and $\overline{X}_e$ while 
$\vec{\tau}$ are Pauli matrices that act on the color space.
The $U(2)$ symmetry is reduced to $U_B(1)\times U_\chi(1)$ in the presence 
of a chemical potential:
\begin{equation}
\begin{array}{ll}
U_B(1): & \quad\quad X_o \rightarrow \exp(i\sigma_3 \phi) X_o, \quad 
\overline{X}_e \rightarrow \overline{X}_e \exp(-i\sigma_3\phi) \cr
U_\chi(1): & \quad\quad X_o \rightarrow \exp(i\phi) X_o,\quad 
\overline{X}_e \rightarrow \overline{X}_e \exp(-i\phi).
\end{array}
\end{equation}
Here $U_B(1)$ is the baryon number symmetry 
$\chi(x)\rightarrow\mathrm{e}^{i\phi}\chi(x),\quad\overline{\chi}(x)
\rightarrow\overline{\chi}(x)\mathrm{e}^{-i\phi}$ and
$U_\chi(1)$ is the chiral symmetry of staggered 
fermions $\chi(x)\rightarrow \mathrm{e}^{i\phi\varepsilon(x)}\chi(x)$, 
$\overline{\chi}(x)\rightarrow\overline{\chi}(x)
\mathrm{e}^{i\phi\varepsilon(x)}$ 
where $\varepsilon(x)=(-1)^{x_{t}+x_{1}+x_{2}+...+x_{d}}$.

\section{DIMER-BARYONLOOP REPRESENTATION}

\begin{figure}
\begin{center}
\includegraphics[width=0.5\textwidth]{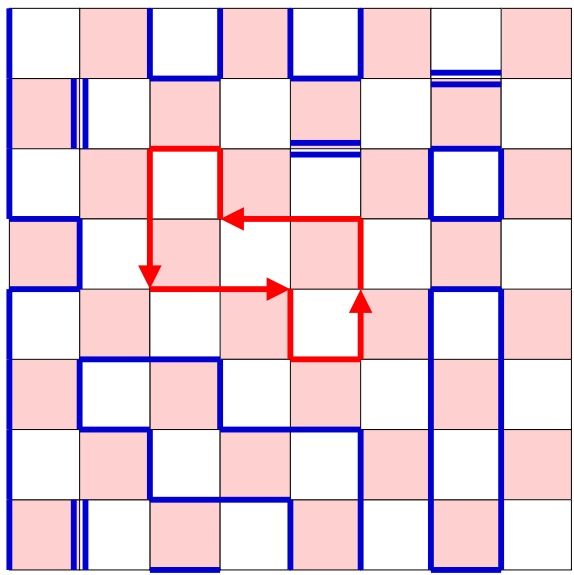}
\end{center}
\caption{\label{fig0} An example of a dimer-baryonloop configuration.}
\end{figure}

One of the computational advantages of the strong coupling limit is that
in this limit it is possible to rewrite the partition function,
\begin{equation}
Z = \int [DU][d\overline\chi d\chi] \exp(-S),
\end{equation}
as a sum over configurations containing gauge invariant objects 
\cite{Rossi:1984cv,Wolff:1984we,Karsch:1988zx}. In our case these objects 
turn out to be dimers and baryonloops. A lattice configuration $K$
of dimers and baryonloops is constructed as follows: 
\begin{itemize}
\item[(a)] Every link of the lattice connecting the site $x$ with the 
neighboring site 
$x+\hat{\alpha}$ contains either a dimer $k_\alpha(x)=0,1,2$ or a directed 
baryonic bond $b_\alpha(x)=-1,0,1$. $b_\alpha(x) = 1$ indicates the direction 
is from $x$ to $x+\hat{\alpha}$ and $-1$ implies it is from $x+\hat{\alpha}$ 
to $x$. $k_\alpha(x)=0$ and $b_\alpha(x)=0$ means that the link does 
not contain any dimer or baryonic bond. In our notation we also allow 
$\hat{\alpha}$ to be negative. Thus, if $\alpha$ was positive, 
$k_{-\alpha}(x)$ and $b_{-\alpha}(x)$ will represent dimers and baryonic 
bonds connecting $x$ with $x-\hat{\alpha}$.
\item[(b)] If a site is connected to a baryonic bond then it must have 
exactly one incoming baryonic bond and one outgoing baryonic bond. Thus 
baryonic bonds always form self-avoiding baryonloops.
\item[(c)] Every lattice site $x$ that does not contain a baryonic bond 
must satisfy the constraint 
\[
\sum_\alpha k_\alpha(x) = 2
\] 
where the sum includes negative values of $\alpha$. 
\end{itemize}
An example of a dimer-baryonloop configuration is shown in Figure \ref{fig0}.
Note that sites connected by $k_\alpha(x)=1$ also form loops. Given the 
set $\{K\}$ of such dimer-baryonloop configurations the partition 
function of the theory described by eq.(\ref{scact}) can be rewritten as 
\cite{Klatke:1989xy},
\begin{equation}
Z=\sum_{\{ K\}} \quad \Bigg\{
\prod_x \quad T^{[k_t(x) + |b_t(x)|]}\exp[2\mu a_t b_t(x)]\Bigg\}
\label{dblpf}
\end{equation}
where $T=\frac{1}{a_{t}^{2}}$. Note that the partition function has
been written as a statistical mechanics of dimers and baryonloops 
with positive definite Boltzmann weights. It is possible
to extend the Monte Carlo algorithm developed in \cite{Adams:2003cc}
and apply it to this problem. The details of the algorithm will 
be published elsewhere.

\section{OBSERVABLES}

A variety of observables can be measured with our new algorithm. We will 
focus on the following:

\begin{itemize}

\item[(a)] The chiral two point function, given by
\begin{equation}
G_C(z,z') = \Bigg \langle \overline{\chi}(z)\chi(z) \ 
\overline{\chi}(z')\chi(z')
\Bigg\rangle
\end{equation}
and the chiral susceptibility,
\begin{equation}
\chi_{C}\equiv \frac{1}{\Omega}\sum_{z'} G_C(z,z')
\end{equation}
where $\Omega$ is the lattice space-time volume.

\item[(b)] The diquark two point function, given by
\begin{equation}
G_B(z,z') = 
\Bigg \langle 
\chi_1(z)\chi_2(z) \  \overline{\chi}_2(z')\overline{\chi}_1(z')
\Bigg\rangle
\end{equation}
and the diquark susceptibility,
\begin{equation}
\chi_{B}\equiv \frac{1}{\Omega}\sum_{z'}  G_B(z,z')
\end{equation}

\item[(c)] Baryon density, defined as
\begin{equation}
n_{B}\equiv\frac{1}{2\Omega}\frac{\partial \ln Z}{\partial\mu}
\end{equation}

\item[(d)] The helicity modulus associated with the $U(1)$ 
chiral symmetry, which we define as
\begin{equation}
Y_{C} \equiv
\frac{1}{d \Omega_s}\sum_{\alpha=1,2,..,d}
\Bigg\langle\Big(\sum_{x}A_{\alpha}(x)\Big)^{2}\Bigg\rangle
\label{yc}
\end{equation}
where
\begin{equation}
A_{\alpha}(x)= \varepsilon(x) \left[ |b_\alpha(x)| + k_\alpha(x)\right]
\end{equation}
and $\Omega_s$ is the spatial lattice volume.

\item[(e)] The helicity modulus associated with the $U(1)$ 
baryon number symmetry, which we define as
\begin{equation}
Y_{B}\equiv \frac{1}{d \Omega_s}\sum_{\alpha=1,2,..,d}
\Bigg\langle \Big(\sum_{x}B_{\alpha}(x)\Big)^{2}\Bigg\rangle
\label{yb}
\end{equation}
where
\begin{equation}
B_{\alpha}(x)= \left[ b_\alpha(x) \right]
\end{equation}

\end{itemize}
Both $Y_C$ and $Y_B$ are diagonal observables and can be calculated
configuration by configuration in the dimer-baryonloop language and 
averaged. On the other hand $G_C(z,z')$ and $G_B(z,z')$ are examples
of off-diagonal observables and can be measured by exploiting the 
special properties of the directed loop update \cite{Adams:2003cc}.

\section{EXPECTED PHASE DIAGRAM}

At $\mu=0$, one expects a finite temperature phase transition separating 
the low temperature ($U(2)\rightarrow U(1)$) broken phase and the high 
temperature symmetric phase. 
Similarly, for small $\mu$ there is a phase transition separating the 
low temperature phase where $U_B(1)\otimes U_\chi(1)$ is broken completely 
and the high temperature phase which is symmetric. The order of both these
transitions remain unclear. At zero temperature as $\mu$ increases one 
expects the lattice to get saturated with baryons which leads to
a phase transition from a super-fluid to a normal phase. Renormalization 
group arguments suggests that if this phase transition is second order it 
will be a mean field transition\cite{Fisher89}.

Little is known about the phase diagram in two spatial dimensions. One
possible phase structure was discussed in \cite{Dunne:2003ji} in the 
context of the continuum theory using an effective field theory approach.
However, due to infrared divergences that occur at finite temperatures
a complete picture could not be inferred. Although continuous symmetries 
cannot break in two spatial dimensions at finite temperatures, a
Berezinky-Kosterlitz-Thouless (BKT) type phase transition can exist 
\cite{BKT}. 

\section{RESULTS}
 
Although spontaneous symmetry breaking cannot occur in a finite volume,
one can still conclude that the symmetry is broken in the infinite volume 
limit by studying the behavior of various observables as a function of 
the volume. In our case the $U(2)$ symmetry at $\mu=0$  implies 
$G_C(z,z') = 2 G_B(z,z')$ and so $\chi_C = 2\chi_B$. In the presence of
a chemical potential since the $U(2)$ symmetry is broken, this equality
no longer holds. The formation of a diquark condensate can be inferred from 
the growth of $\chi_B$ with the volume. Further, the helicity modulus $Y_C$ 
and $Y_B$, both must reach a non-zero constant if the $U_\chi(1)$ and 
$U_B(1)$ are broken. All these expectations can be understood quantitatively 
using chiral perturbation theory in the $\epsilon$-regime based on an 
effective action, which at $\mu=0$ turns out to be
\begin{equation}
S_{\rm eff} = \int d^dx \quad \Bigg[
\frac{F^{2}}{2}(\partial_{\mu}\vec{u})\cdot (\partial_{\mu}\vec{u})
+\frac{B^{2}}{2}(\partial_{\mu}\vec{S})\cdot(\partial_{\mu}\vec{S})\Bigg],
\label{eft}
\end{equation}
where $\vec{u}(x)$ is a unit two-vector field and $\vec{S}(x)$ is a unit three 
vector field. This chiral Lagrangian is equivalent to other chiral 
Lagrangians found in the literature \cite{Kogut:2003ju}. However, we note 
that the fact that $F$ and $B$ may not be the same was not considered in 
earlier work . Using the chiral Lagrangian is straight forward to extend 
the results of \cite{Hasenfratz:1989pk} to obtain a finite size scaling 
formula for various quantities. 

\subsection{$d=3$ with $\mu=0$}
\label{zcp}

\begin{figure}
\vskip0.3in
\begin{center}
\includegraphics[width=0.75\textwidth]{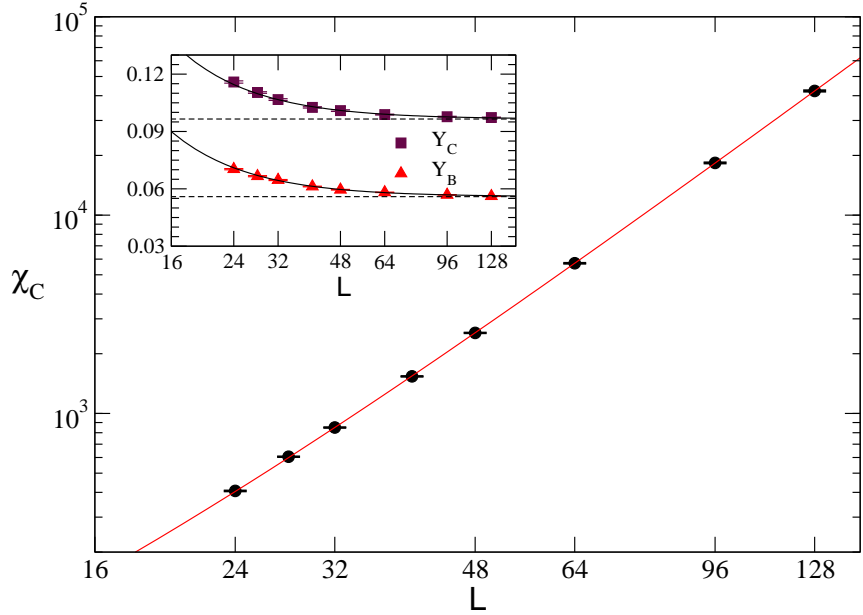}
\end{center}
\caption{\label{fig1} The inset shows $Y_C$ and $Y_B$ as a 
function of $L$ at $T=2.918$. We find $F^2=0.0965(5)$ and $B^2=0.0559(4)$.
Using these values a fit of $\chi_C$ as a function of $L$
gives $\Sigma=0.2372(3)$ and $a=2.11(6)$ with a $\chi^2/DOF = 0.5$.
The main plot shows $\chi_C$ versus $L$ and the fit (solid line).}
\end{figure}

The finite temperature phase transition that restores $U(2)\rightarrow U(1)$ 
symmetry breaking can be studied in our model by tuning $T$ at fixed $L_t$.
We have performed extensive calculations at a fixed $L_t=4$ for different 
spatial lattice sizes $L$ varying from $16$ to $256$ and for many different 
values of $T$. We look for two signatures of the broken phase: 
\begin{itemize}
\item[(a)] Both $Y_C$ and $Y_B$ must go to non-zero constants at large $L$. 
These constants are equal to the low energy constants, $F^2$ and $B^2$ 
in eq.(\ref{eft}), of a three dimensional low energy effective theory. 
We use the relations
\begin{equation}
Y_C = F^2 + b/L + c/L^2; \qquad Y_B = B^2 + b'/L + c'/L^2.
\label{fss2}
\end{equation}
to extrapolate our data to extract $F^2$ and $B^2$.
\item[(b)] The finite size scaling of the chiral susceptibility $\chi_C$ 
can be shown to be
\begin{equation}
\chi_C = \frac{\Sigma^{2}}{3}\lbrace L^3 + 
\beta_{1}(\frac{2}{F^{2}}+\frac{1}{B^{2}})L^2\rbrace + aL
\label{fss1}
\end{equation}
where $\beta_1 = 0.226$ is the shape coefficient for cubic boxes.
\end{itemize} 
Figure \ref{fig1} shows our results at $T=2.918$, a point in the broken phase.
As can be seen from the graph, the above expectations are satisfied 
extremely well. In particular we find $F^2 \neq B^2$.

\begin{figure}
\vskip0.3in
\begin{center}
\includegraphics[width=0.7\textwidth]{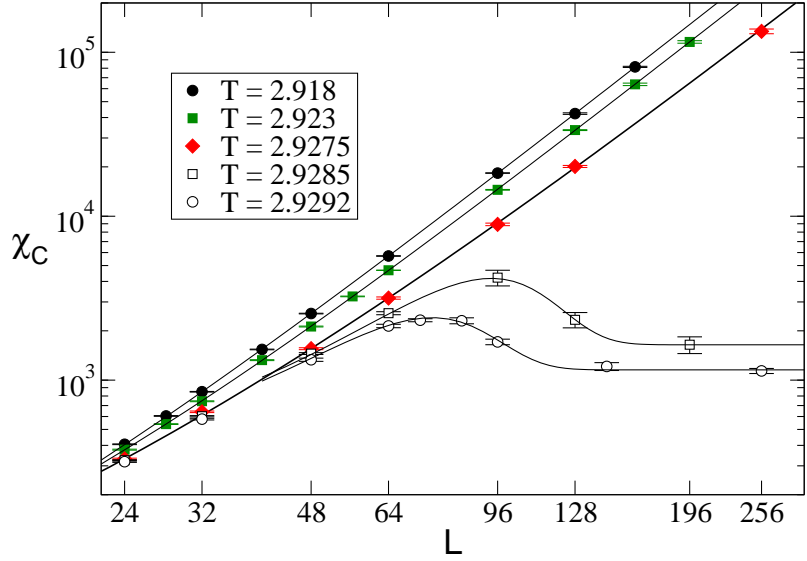}
\end{center}
\caption{\label{fig2} Plot of $\chi_C$ versus $L$ for 
different values of $T$ across the phase transition. The solid lines
at $T=2.9285$ and $T=2.9292$ are fits to the first order form discussed
in the text.}
\end{figure}

Figure \ref{fig2} shows the dependence of $\chi_C$ as a function of $L$ for 
different temperatures. While $\chi_C$ increases as $L^3$ at $T=2.9275$, 
it saturates at $T=2.9285$ and $2.9292$ for large $L$. Note that
extremely large lattices are necessary to see this saturation. Thus, there
is a phase transition such that $2.9275 < T_c < 2.9285$. The 
peculiar non-monotonic behavior of $\chi_C$ before it saturates appears 
to rule out a second order behavior but is consistent with a first order 
transition. We can fit $\chi_C$ to the form
\begin{equation}
\chi_C = \frac{A + B L^3 \exp(-\Delta F L^3)}{1 + C\exp(-\Delta F L^3)}
\label{fss3}
\end{equation}
which can be motivated as arising due to the presence of two phases whose 
free energy densities differ by $\Delta F$. We find this form captures the 
structure in the data well for $L\geq 48$ for both $T=2.9285$ and $2.9292$. 

In the high temperature phase one can compute screening lengths $\xi$
by looking at the exponential decay of $G_C(z,z')$ for large spatial 
separations between $z$ and $z'$. Similarly in the low temperature 
phase $F^2$ and $B^2$ have dimensions of inverse length and hence
can be used to provide natural length scales in the problem. At the 
phase transition we find that none of these length scales diverge 
but all are of the order of $40$ to $50$ lattice units indicating 
that the transition is a rather weak first order transition.

A renormalization group analysis of the fluctuations of the order parameter 
field, which in our case is a complex three vector field $\Phi_i(x),i=1,2,3$,
has been performed using the $\epsilon$-expansion \cite{Jones76,Kawamura88}. 
Interestingly, the universal field theory one studies also 
describes the possible normal-to-planar super-fluid transition in $^3$He 
\cite{Prato}. One finds no stable fixed point which implies that any observed 
phase transition must be a fluctuation driven first order transition. Our
observations favor this conclusion. On the other hand recently it has been
proposed that the $\epsilon$-expansion results may be misleading \cite{Prato}. 
It is well known that first order transitions can arise due to model dependent
features. In order to minimize such dependences it may be useful to 
repeat the above calculation for other values of $L_t$.

\subsection{$d=3$ at zero temperature}

\begin{figure}
\vskip0.3in
\begin{center}
\includegraphics[width=0.7\textwidth]{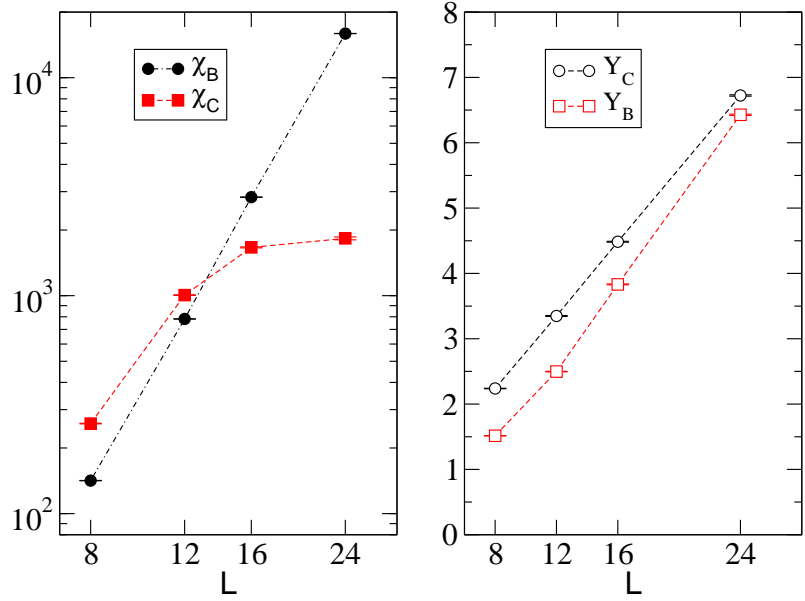}
\end{center}
\caption{\label{fig4} The left plot shows $\chi_B$ and $\chi_C$ as a function 
of $L$ at $\mu=0.01$. As expected $\chi_B$ grows as $L^4$ while $\chi_C$
saturates. For the same value of $\mu$ the right plot shows both $Y_C$ and 
$Y_B$ grow linearly with $L$.}
\end{figure}

In order to study zero temperature results we set $L_t=L$, $T=1$ and
compute observables for various values of $\mu$ and $L$. In this case 
since $U_B(1)\otimes U_\chi(1)$ is expected to be broken completely 
the following should be observed:
\begin{itemize}
\item[(a)] The diquark susceptibility $\chi_B$ should grow with the volume,
\begin{equation}
\chi_B \sim \frac{\Delta^2}{2} L^4
\end{equation}
where $\Delta = \langle \chi_1\chi_2 \rangle = 
\langle \overline{\chi}_2 \overline{\chi}_1\rangle \neq 0$ is the diquark 
condensate.
\item[(b)] The chiral susceptibility $\chi_C$ should saturate with $L$ showing 
that the $G_C(z,z')$ decays exponentially for large separations between
$z$ and $z'$. Note that the two correlators $G_C$ and $G_B$ are no longer 
related by a symmetry when $\mu \neq 0$.
\item[(c)] Both the helicity modulus $Y_C$ and $Y_B$, as defined in
eqs.(\ref{yc}),(\ref{yb}) should grow linearly with $L$.
\end{itemize}
These expectations are borne out in our calculations as can be seen
in Figure \ref{fig4}.

\begin{figure}
\vskip0.3in
\begin{center}
\includegraphics[width=0.7\textwidth]{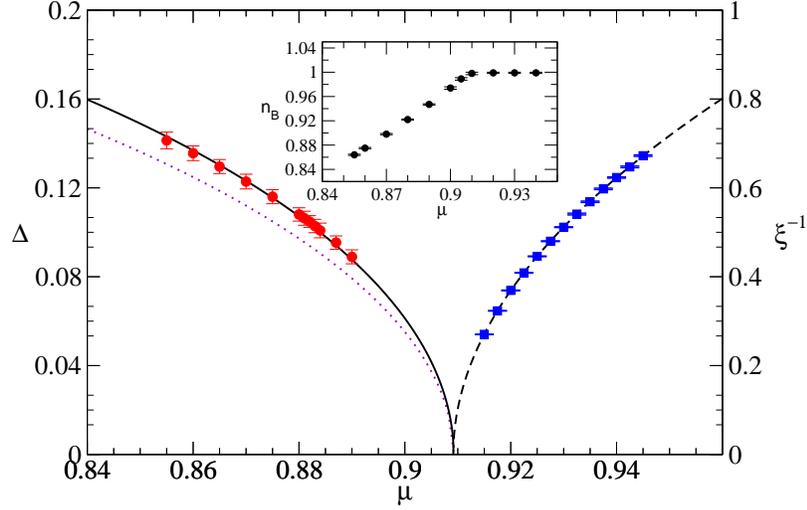}
\end{center}
\caption{\label{fig5} The plot of $\Delta$ and
$\xi^{-1}$ as a function of $\mu$. The dotted line is the mean field
result and the solid line includes the one loop corrections. The 
dashed line is a fit to the form $\xi^{-1} = A \sqrt{(\mu-\mu_c)}$. 
The inset shows the baryon density $n_B$ as a function of $\mu$.}
\end{figure}

As the chemical potential increases the lattice is filled with baryons.
This can be seen in the inset of figure \ref{fig5} where the baryon 
density is plotted as a function of $\mu$. Baryon super-fluidity cannot 
exist beyond saturation and this leads to a phase transition to the 
non-super-fluid phase. Renormalization group arguments show that this 
phase transition must belong to the mean field universality class 
for $d\geq 2$ \cite{Fisher89}. It was recently shown that the critical 
chemical potential $\mu_c = 0.5\cosh^{-1}(\sqrt{10}) = 0.909223..$ in 
the mean field approximation \cite{Nishida:2003uj}. The diquark 
condensate was also shown to be\footnote{There is a factor
of two mismatch the formula quoted here and what can be found 
in \cite{Nishida:2003uj}. The origin of this mismatch is the
normalization of our kinetic term in eq.(\ref{scact}) as compared
to the one in \cite{Nishida:2003uj}.}
\begin{equation}
\Delta = \sqrt{\frac{1}{18}\left(\sqrt{10}-\cosh(2\mu)\right)}
\label{mfdelta}
\end{equation}
In our calculations we extracted $\Delta$ by fitting $\chi_B$ to 
the relation
\begin{equation}
\chi_B = \frac{\Delta^2}{2}[L^4 + A L^2 + B].
\end{equation}
Figure \ref{fig5} shows our results along with the mean field result 
and the result with one-loop corrections. We find that $\mu_c$
is in excellent agreement with mean field theory while $\Delta$ 
requires the inclusion of one-loop corrections.

For $\mu > \mu_c$ it costs energy to remove a single baryon and 
we expect this energy to grow as $(\mu-\mu_c)$. Since this 
phase describes non-relativistic particles the spatial correlation
length $\xi$, obtained from $G_B(z,z')$, must scale as 
$1/\sqrt{(\mu-\mu_c)}$. Figure \ref{fig5}, also shows that this
expectation is borne out. 

\subsection{$d=2$ Phase Diagram}

We have also studied both the zero temperature and finite temperature phase 
transitions in two spatial dimensions. While the zero temperature 
transition is a mean field transition from a super-fluid phase to the 
saturated phase like in $d=3$, the finite temperature phase transition 
at $\mu=0$ is different and interesting. Mermin-Wagner theorem forbids 
the breaking of a continuous symmetry in $d=2$ at finite temperatures. 
However, the $U(1)$ part(s) of the $U(2)$ symmetry at $\mu=0$ can
undergo a BKT type phase transition. This can result in long range 
correlations in $\chi_C$ and $\chi_B$ at low temperatures. If this
is true one would expect
\begin{equation}
\lim_{L\rightarrow \infty} Y_C = 
\left\{\begin{array}{cc} 
\mathrm{Const.} & T < T_c \cr
0 &  T > T_c
\end{array} \right.
\end{equation}
while $\lim_{L\rightarrow \infty} Y_B = 0$. Further, the screening lengths 
$\xi$ obtained from $G_C(z,z')$ will behave as
\begin{equation}
\xi = A \exp\Bigg(\frac{B}{\sqrt{T-T_c}}\Bigg), \quad T > T_c
\end{equation}
\begin{figure}
\vskip0.3in
\begin{center}
\includegraphics[width=0.75\textwidth]{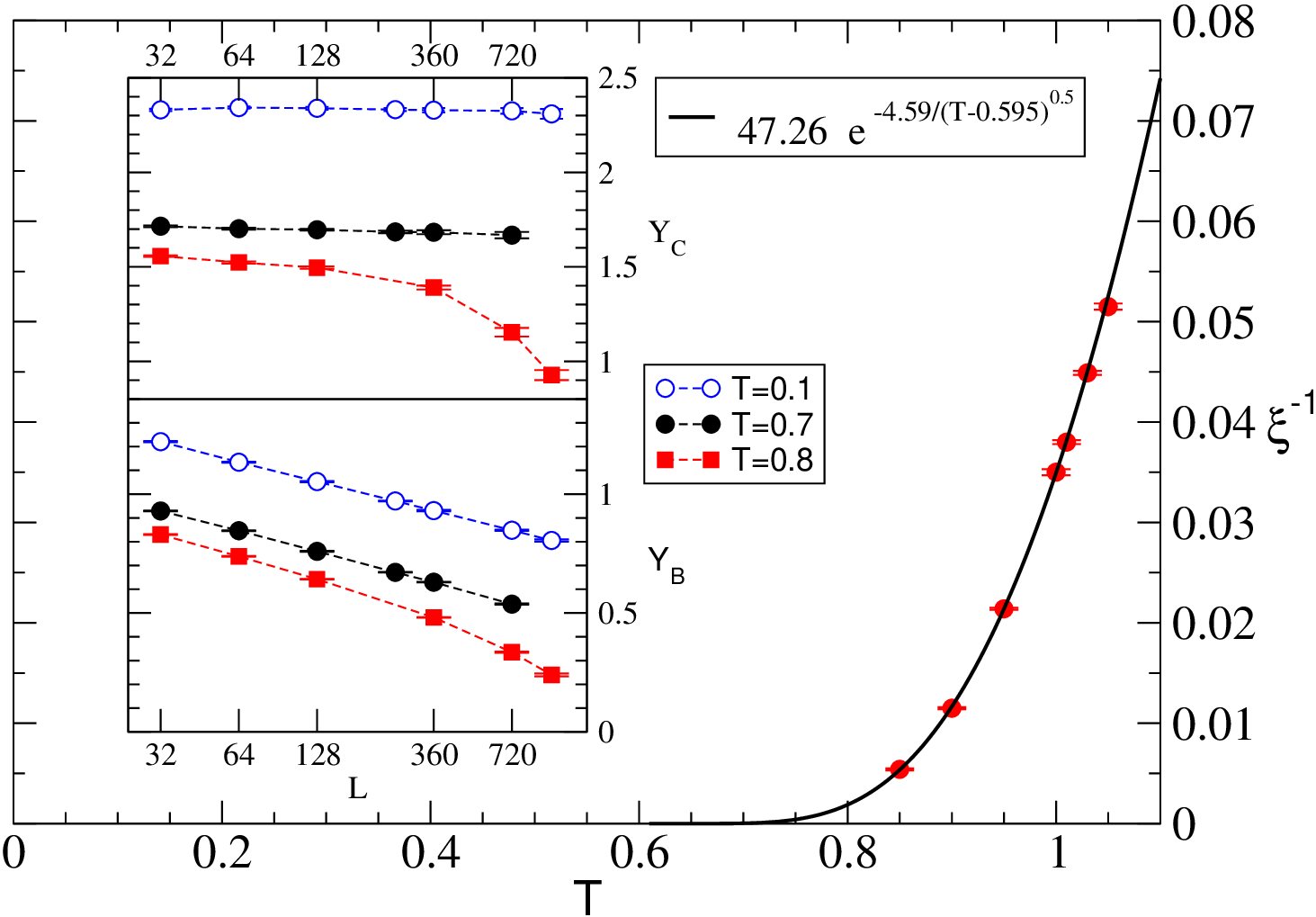}
\end{center}
\caption{\label{fig6} }
\end{figure}

Figure \ref{fig6} shows that indeed our data is consistent with these
expectations. We see that $Y_C$ is a constant and $Y_B$ decreases (although
very slowly) with $L$ at $T=0.1$. For $T \geq 0.7$ we can begin to see
both $Y_C$ and $Y_B$ decrease with $L$. For $0.8 \leq T \leq 1.03$  
$\xi$ extracted from $G_C(z,z')$ fits well to the BKT form with
$T_c \sim 0.6$. In a typical BKT phenomena, the super-fluid density 
is expected to show a universal jump at the transition. A naive estimate 
of this jump suggests that $Y_C(T=T_C) = 2/\pi \sim 0.6366...$, in 
our normalization. Our data shows a different jump suggesting that
the transition is quantitatively different although qualitatively
similar to the familiar BKT transition.

\section*{Acknowledgments}

We thank S. Hands, C. Strouthos, T. Mehen, R. Springer, D. Toublan and 
U.-J. Wiese for helpful comments. This work was supported in part by the 
Department of Energy (DOE) grant DE-FG02-03ER41241. The computations 
were performed on the CHAMP, a computer cluster funded in part by the DOE. 
We also thank Robert G. Brown for technical support and allowing us to use 
his computer cluster for additional computing time.

\end{document}